# First $^{17}$O NMR study of a $^{17}$O enriched LaMnO$_3$ stoichiometric crystal

L. Pinsard Loreynne-Gaudart[1], A. Trokiner[2], S. Verkhovskii[2,3], A. Gerashenko[2,3], N. Dragoe[1]

[1]*Institut de Chimie Moléculaire et des Matériaux d'Orsay, Université Paris Sud, 91405 Orsay Cedex, France*
[2] *LPEM, ESPCI ParisTech, UMR 8213, CNRS, 75005 Paris, France,*
[3] *Institute of Metal Physics, Ural Branch of Russian Academy of Sciences, 620041 Ekaterinburg, Russia*

**Abstract**

We present the synthesis and the NMR characterization of a $^{17}$O enriched LaMnO$_3$ crystalline sample. We checked that it is single phase and, more important, stochiometric in oxygen. Its $^{17}$O enrichment estimated by NMR is about 5.5%. These first $^{17}$O NMR results obtained at T=415K in a undoped parent LaMnO$_3$ manganite demonstrate that the two oxygen sites of the structure probe very different Mn spin correlations in the paramagnetic orbital ordered phase. This work opens the way to study experimentally the interactions responsible for the orbital order.

## I - Introduction

Transition metal oxides remain one of the main topics of solid state physics researches and advanced technological applications. The richness of magnetic and electronic properties of these materials is tightly connected to the orbital degree of freedom of the outer d-electrons of the transition metal ions, e.g., the Jahn-Teller Mn$^{3+}$ ions in the LaMnO$_3$ manganites and depends on the type of occupied orbitals **[1-3]**. An excess of oxygen content, x, yielding LaMnO$_{3+x}$ corresponds to a hole doping which induces a change of magnetic and electronic properties **[4-5].** To achieve an understanding of the spin and orbital Mn correlations behaviour of doped compounds like LaMnO$_{3+x}$ or La$^{+3}_{(1-x)}$Sr$^{2+}_x$MnO$_3$, it is desirable, as a starting basis, to scrutinize these issues in the undoped parent compound, LaMnO$_{3.0}$.

In LaMnO$_3$ perovskite structure, each oxygen is shared by two Mn ions through the Mn(e$_g$)-O(2p) covalent bond yielding a three dimensional network of corner shared MnO$_6$ octahedra. Moreover, the pathway of the Mn-Mn superexchange interactions includes the O(2p) orbitals. Key questions which concern short-range spin correlations of Mn ions and features of the Mn-O chemical bonding, appearing in the orbital ordered paramagnetic (PM) state, can be clarified by $^{17}$O nuclear magnetic resonance (NMR) **[6]**. Indeed, the $^{55}$Mn nucleus is not an appropriate probe for a study of the paramagnetic state since its NMR signal is undetectable above the Néel temperature, T$_N$, due to too short relaxation time. Among the oxygen isotopes only $^{17}$O has non-zero nuclear spin. Its nuclear spin $^{17}I$ = 5/2 probes the Mn electronic spins through nuclear hyperfine interactions. Since the natural abundance of $^{17}$O is too low (0.037%), isotopic enrichment is required for NMR studies in solid compounds with broad spectra.

Not many $^{17}$O enriched crystals have been synthesized before due to sizeable cost and difficulty to grow a high-quality $^{17}$O enriched crystal. A $^{17}$O enriched crystal of Al$_2$O$_3$ has been obtained by floating zone method **[7]** whereas MgO, LaAlO$_3$ **[8]** and CeO$_2$ **[9]** have been obtained by flux growth. On the other hand crystals of La$_{2-x}$Sr$_x$CuO$_4$ **[10],** BaTiO$_3$ and SrTiO$_3$ **[11]** were enriched with $^{17}$O diffusion after the synthesis. This last method is applicable when the enrichment treatment doesn't change the oxygen stoichiometry. In our case, such a procedure should yield an over-stoichiometric oxygen sample LaMnO$_{3+\delta}$ **[12]**. Moreover, performing the isotopic exchange on a single crystal requires very long time and many treatments due to limited diffusion. Another method is to grow a crystal with $^{17}$O enriched starting materials. The disadvantage of this procedure is the reduction of the sample during the growth at high temperature and significant loss of $^{17}$O is expected due to the gas exchange. Therefore we first synthesised a rod with natural isotopic distribution then $^{17}$O exchange was performed on it, before the growth by floating zone method.

In this paper we present a preliminary $^{17}$O NMR study of an undoped LaMnO$_3$ crystal at T=415K in the PM orbital ordered phase. To our knowledge, this is the first $^{17}$O NMR study of LaMnO$_3$. The synthesis of a stoichiometric LaMnO$_3$ crystal enriched with $^{17}$O is described. The obtained crystal was characterised by X-ray and magnetization measurements whereas its $^{17}$O enrichment was estimated by NMR in a crushed part of the crystal. We discuss about the identification of the two oxygen sites, which probe respectively distinct spin correlations of the Mn-neighbours in the orbital ordered phase.

## II – Experimental

Crystal growth was carried out using a floating-zone furnace equipped with double hemi-ellipsoidal mirrors (NEC, SC-N15HD) in which two halogen lamps were used as heating sources. The crystal was grown at a rate of 5 mm/h under flowing argon. These conditions lead to obtain stoichiometric single crystal of LaMnO$_3$ **[13].**

Magnetic properties were measured with a SQUID magnetometer (Quantum Design) from 4K to 300K in a magnetic field $H = 50$ Oe.

The NMR measurements were performed at T=415K with an AVANCE III BRUKER spectrometer operating at 11.747 T. At this field, the Larmor frequencies of $^{17}$O and $^{139}$La nuclei are close, 67.8 and 70.647 MHz respectively. As the quadrupolar interaction is present for $^{17}$O and $^{139}$La nuclei ($^{139}I$=7/2), both spectra are very broad, approximately 2 and 25 MHz, respectively. Thus, a method of frequency sweeping was used. A pulse sequence $\alpha - \tau - 2\alpha - \tau -$ (echo) was used with delays $\tau = 12$ and 90µs and a pulse duration $\alpha$=1.1µs, shorter than the one which optimizes the echo signal amplitude of both nuclei. The total spectra were obtained by summing the Fourier-transformed half-echo signals acquired at equidistant operating frequencies. For instance, 116 echoes were acquired every 0.1 MHz from 60.5 to 83.5MHz. The spin-spin relaxation times, $T_2$, were determined by varying $\tau$. When estimating the $^{17}$O enrichment, the spin-echo amplitudes were extrapolated back to $\tau$=0 and the signal was divided by $\nu^2$, proportional to the NMR sensitivity. The repetition time was chosen 5 times larger than the spin-lattice relaxation time of both nuclei. With all these precautions, the intensity of the NMR lines is proportional to the number of nuclei.

## III – Synthesis and characterisation of the $^{17}$O enriched crystal

The starting materials for preparation of a feed rod were high purity powders of La$_2$O$_3$ and MnO$_2$. La$_2$O$_3$ powder was previously heated at 900°C in order to remove the adsorbed water and an excess of 5% wt MnO$_2$ was added to compensate for the volatilization of manganese during the single crystal growth. The powders were thoroughly mixed and fired in air at 1200°C for 24 h. The resulting powder was compacted in the form of a rod of about 3.5 mm diameter under a hydrostatic pressure of 2.5 kbar. The rod was sintered at 1300–1400°C for 48 h. It was then isotopically enriched at high temperature by using an exchange with $^{17}$O$_2$ gas (74% $^{17}$O enrichment). The sintered sample, 6g, was placed in a platinum crucible in a quartz tube placed in a furnace. The quartz tube was filled with a slightly over one bar of $^{17}$O$_2$ at room temperature. The temperature was then increased to 950 °C for 7 days. This procedure was repeated three times with refilling of a fresh $^{17}$O gas, in order to obtain acceptable $^{17}$O enrichment.

The isotope ratio $^{17}$O/$^{16}$O is normally estimated from the weight changes of the samples. This estimation was difficult for different reasons: (i) we don't know the exact composition of the rod due to the excess of manganese oxide which was added to compensate for the evaporation observed during the single crystal growth, (ii) in our experimental conditions, the obtained rod presents an excess of oxygen. A rough estimation of the $^{17}$O content in the feed rod gave about 20%-25%.

The obtained crystal was checked by optical microscopy in polarized light and twins were observed, as expected for this compound **[14].**

Powder X-ray diffraction experiments were performed on an X-ray diffractometer PANalytical X'Pert System equipped with an X'Celerator detector and using Cu K$_{\alpha1}$ radiation, on a crushed piece of crystal. Rietveld analysis **(fig. 1)** showed that a single phase was obtained with a P*bnm* orthorhombic space group (a=5.5379(1)Å, b=5.7484(1)Å, c=7.6950(1)Å). As noted in ref. **12,** the lattice parameters for LaMnO$_3$ are very sensitive to the amount of Mn$^{4+}$ and therefore to the oxygen non-stoichiometry. The obtained values are close to those obtained for stoichiometric composition.

The magnetization data of a slice of the $^{17}$O enriched LaMnO$_3$ crystal **(fig. 2)** show that the onset of the antiferromagnetic phase occurs at T$_N$ ~ 139K. T$_N$ is also sensitive to the stoichiometry of the sample. As the value of 139K is very close to the one found in natural undoped single crystals **[15],** it confirms that the $^{17}$O enriched LaMnO$_3$ crystal is stoichiometric, in agreement with X-ray results.

## IV – $^{17}$O NMR study of LaMnO$_3$

Two $^{17}$O spectra are expected since in the orthorhombic structure of LaMnO$_3$ there are two oxygen sites, O1 and O2 in the ratio 1:2. The O1 site also named apical is located between the *ab*-planes, its two first Mn neighbors are antiferromagnetically (AF) correlated whereas O2 site, the equatorial site, is in the *ab*-planes between two ferromagnetically (FM) correlated Mn first neighbors (inset of fig. 1). The spin correlations are long range in the AF state and short range in the PM state. Furthermore, due to the orbital ordering, the Mn and O orbital's involved in the Mn-O bonds are different for both sites.

As explained before the $^{17}$O and $^{139}$La NMR spectra overlap. The quadrupolar interaction yields five NMR lines for each $^{17}$O site and 7 lines for $^{139}$La if twinning was absent. In total, more than a half hundred NMR lines are expected in the crystal due to the possible six twins of the orthorhombic LaMnO$_3$ **[16].** In order to distinguish both nuclei and to identify O1 and O2 sites, we have studied a crushed part of the crystal to obtain a powder pattern. In addition, we have compared the NMR signal of two samples, a natural stoichiometric LaMnO$_{3.0}$ (LaMn$^{16}$O$_3$) synthesized in the same conditions than those studied in ref. 15 and the $^{17}$O enriched (LaMn$^{17}$O$_3$) sample. For the latter, the $^{139}$La nuclei, with 100% natural abundance, are responsible of the main part of the NMR spectra.

A part of the NMR spectrum of LaMn$^{16}$O$_3$ and LaMn$^{17}$O$_3$ samples are displayed on **fig. 3** and the whole spectrum of the former is shown in the inset. Both spectra were acquired with the delay τ=12µs. The intensity of the spectra can be directly compared since the mass of both samples is the very same. For LaMn$^{16}$O$_3$ the spectrum is only due to $^{139}$La nuclei since it is the only NMR signal existing in this frequency range. The LaMn$^{17}$O$_3$ spectrum shows two additional lines at 68.5 and 69.4 MHz which we attribute to the expected $^{17}$O lines. The corresponding shift of the maximum of each line is K$_{max}$= 0.98±0.04% and 2.35±0.06%, respectively. The $^{17}$O satellite lines are hardly observed, in contrast to $^{139}$La NMR. This shows that for both O sites, the magnetic interaction overcomes the quadrupolar one. Let us consider the line shifts. The local field is expected to be larger at the equatorial sites since it is produced by the two FM correlated Mn first neighbors whereas the two Mn first neighbors of the apical sites are AF correlated. This enables to attribute the more shifted line to O2 and the less shifted one to O1. In **fig 4** are displayed two LaMn$^{17}$O$_3$ spectra acquired with different echo delays showing that the two $^{17}$O spectra have very different spin-spin relaxation times, $^{17}$T$_2$, since for τ=90µs, only O1 line remains. In LaMn$^{16}$O$_3$ the decay of the echo amplitude of the central line is exponential with $^{139}$T$_2$ = 360 ± 15µs. In contrast, at the position of O1 and O2 lines, a good fit of the echo decay requires two exponentials as $^{17}$O and $^{139}$La nuclei contribute to the signal at these frequencies. At O2 frequency, the decomposition yields 360 ± 15µs and 32 ± 3µs corresponding respectively to $^{139}$T$_2$ and $^{17}$T$_2$(O2). At O1 frequency, the two T$_2$ components are very close so that we fixed the first one to $^{139}T_2=360 \pm 15\mu s$ and found for the second $^{17}T_2(O1)=340 \pm 25\mu s$. We find the ratios: $^{17}T_2(O1)/\,^{17}T_2(O2) \approx 11$ and $[(h_{loc,\,static}(O2)/h_{loc,\,static}(O1)]^2 \approx 6$. In the case of uncorrelated fluctuations of the Mn spins one expects that $T_2(O1)/T_2(O2) = [(h_{loc,static}(O2)/\,h_{loc,\,static}(O1)]^2$ **[Ref. 17].** The too large value of the $T_2$-ratio is explained by the spin correlations of the Mn-neighbors which exist in the orbital ordered PM phase.

The $^{17}O$ NMR spectrum acquired with $\tau=12\mu s$ demonstrates that both oxygen sites of the structure have been $^{17}O$ enriched. The decomposition into O1 and O2 spectra is shown on the inset of **fig 4**. It was obtained with linear combinations of the spectra of fig. 4 and the LaMn$^{16}$O$_3$ spectrum. We have estimated $c_1$ and $c_2$, the $^{17}O$ enrichment of O1 and O2 site, respectively. For that, we have compared $^{139}$La and $^{17}$O NMR signal intensities extrapolated to $\tau=0$ in the natural and $^{17}$O enriched samples, respectively. We find $c_1 = c_2 = 5.5 \pm 1\%$ when assuming that the relative proportion of O1 and O2 sites is 1:2 as indicated by the structure. Within the accuracy of our estimation the $^{17}O$ enrichment is the same for both sites. This estimation supports the assignment of the two $^{17}O$ NMR spectra.

## V Conclusion

We have synthesized a $^{17}O$ enriched LaMnO$_3$ crystalline sample in an optical floating-zone furnace. The characterization of the obtained single crystal showed that LaMnO$_3$ has a single phase orthorhombic structure and good crystallinity. Moreover, the comparison with previous results shows that the LaMn$^{17}$O$_3$ crystal is stoichiometric in oxygen.

The $^{17}O$ enrichment of the feed rod, 20-25%, was estimated from the weight changes of the sample. In the crystal, the $^{17}O$ enrichment estimated by NMR is about 5.5%. The drawback of $^{17}O$ loss during the synthesis is largely compensated by the stoichiometric composition of the crystal. As already mentioned, it is desirable to have a careful understanding of the undoped parent compound LaMnO$_{3.0}$, as a starting basis in order to study the evolution of the Mn-Mn spin correlations probed by oxygen nuclei in O1 and O2 sites in doped compounds. In addition, the two oxygen sites of the structure have the same $^{17}O$ enrichment. This method of synthesis of $^{17}O$ enriched stoichiometric LaMnO$_3$ crystal can be recommended for other refractory oxides.

This first $^{17}O$ NMR study in LaMnO$_3$ was performed on a crushed part of the crystal to assign unambiguously the $^{17}O$ and $^{139}$La NMR signals, thus avoiding the complexity of the spectrum in the twinned crystal with overlapping $^{17}O$ and $^{139}$La signals. The two $^{17}O$ spectra could be attributed to the apical (O1) and equatorial (O2) sites. The static and fluctuating components of the local field produced by the Mn$^{3+}$ ions are larger for the equatorial oxygen than for the apical one. This is a direct consequence of the spin correlations of the Mn-neighbors which exist in the orbital ordered paramagnetic phase. The NMR study of the single crystal in order to probe accurately the charge and spin environments of both oxygen sites in the paramagnetic state up to the transition to the orbital disorder phase is now possible.

**Acknowledgements**
We are grateful to P. Monod for his help with SQUID equipment. We also acknowledge useful discussions with A. Revcolevschi. S.V. and A.G. are grateful to ESPCI and CNRS for hospitality and support.

**FIGURE CAPTIONS**

Figure 1: Observed (red points) and calculated (black line) and difference (blue) powder diffraction patterns of LaMnO$_3$ at 300K ($\lambda$ = 1.5406Å). The expected Bragg peak positions are marked below the profile fit as green vertical bars. Inset: schematic structure of LaMnO$_3$ showing the apical (O1) and the equatorial (O2) oxygen sites.

Figure 2: Field and zero field cooled magnetisation in a slice of the crystal rod (84.5 mg) measured with the magnetic field (50 Oe) parallel to the growing axis.

Figure 3: Part of the NMR spectra acquired at T=415K with $\tau$ =12µs in La Mn$^{16}$O$_3$ and LaMn$^{17}$O$_3$ crushed crystals. Both samples have the same mass (297mg). The Larmor frequencies $^{17}\nu_L$ ($^{17}$O) and $^{139}\nu_L$ ($^{139}$La) are shown by arrows. Inset: whole frequency range of the $^{139}$La NMR spectrum of the natural sample.

Figure 4: Part of the NMR spectrum in LaMn$^{17}$O$_3$ acquired at T=415K with two different $\tau$ delays. The vertical dotted lines indicate the position of the two $^{17}$O NMR lines. The amplitude of the spectrum acquired with $\tau$=90µs is multiplied by 1.65. Inset: Decomposition of the NMR spectra for $^{17}$O nuclei located in O1 and O2 sites. The amplitudes corresponds to the delay $\tau$=12µs.

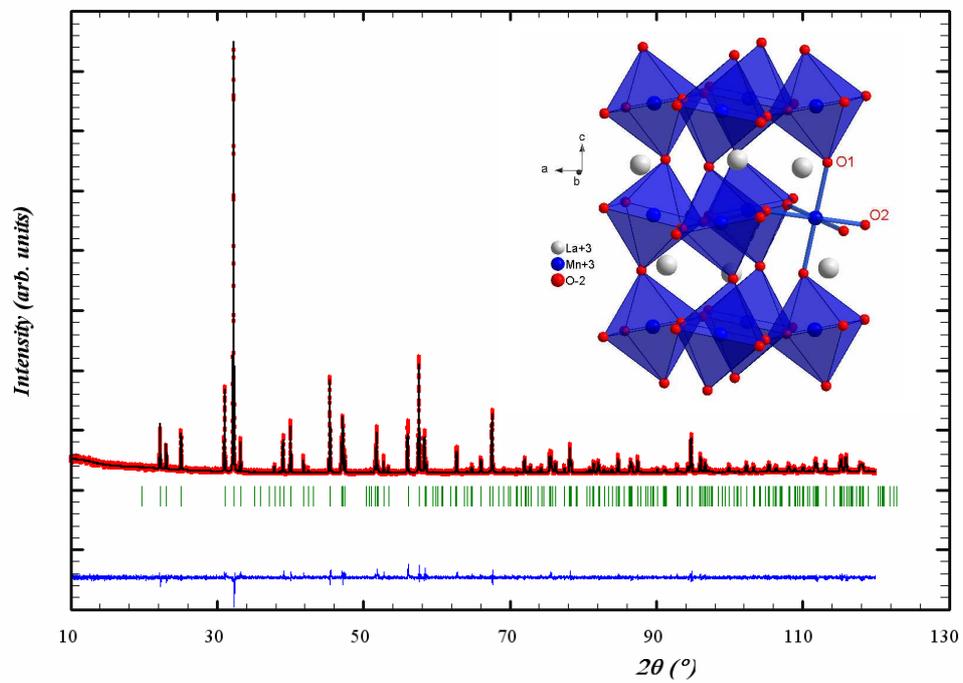

Figure 1

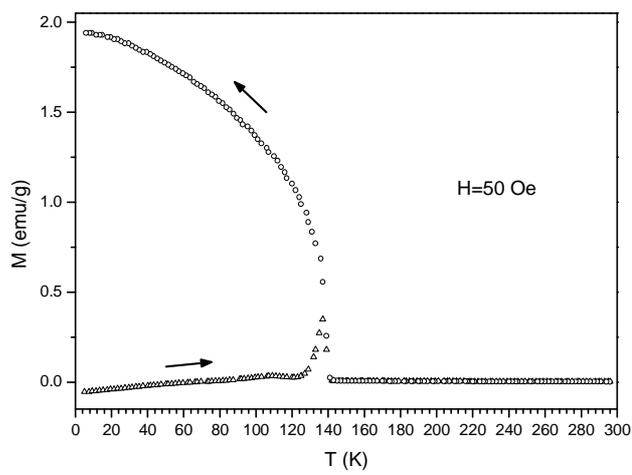

Figure 2

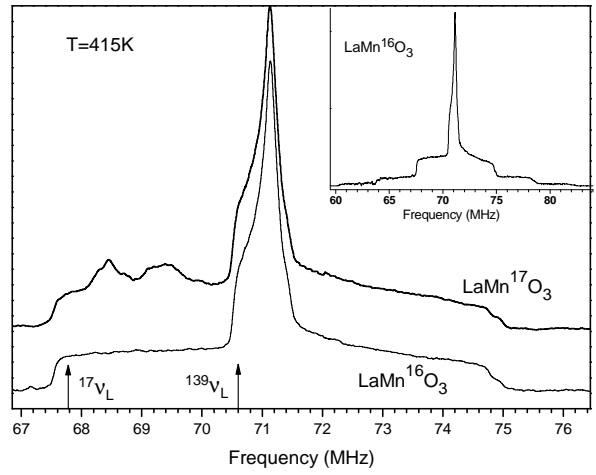

Figure 3

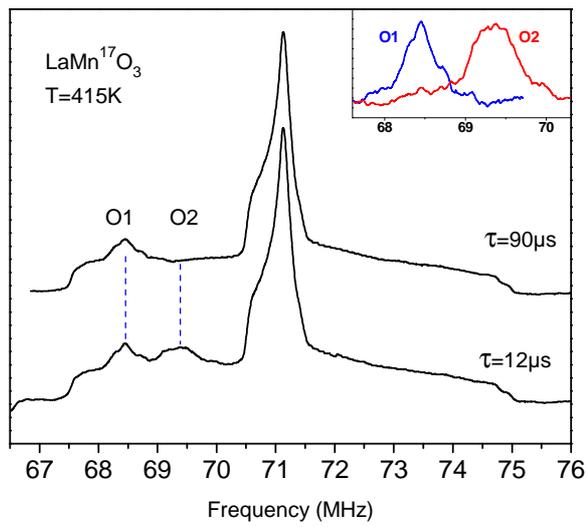

Figure 4